\documentclass[runningheads]{llncs}
\usepackage[T1]{fontenc}
\usepackage{graphicx}
\usepackage[colorlinks = true, allcolors = black]{hyperref}
\usepackage{orcidlink}
\usepackage{subcaption}
\begin{document}

\title{Exploring the Effect of Robotic Embodiment and Empathetic Tone of LLMs on Empathy Elicitation}

\titlerunning{Exploring the Effect of Robotic Embodiment and Empathetic Tone of LLMs}
\author{Liza Darwesh~\orcidlink{009-0004-2874-8122}$^*$\and
Jaspreet Singh~\orcidlink{0009-0007-6893-401X}$^*$\and
Marin Marian~\orcidlink{0009-0005-1039-719X}$^*$ \and
Eduard Alexa~\orcidlink{000-0002-3234-2246}$^*$\and
Koen Hindriks~\orcidlink{0000-0002-5707-5236}\and
Kim Baraka~\orcidlink{000-0003-4381-4234}}
\authorrunning{Darwesh, Singh, Marian, Alexa, et al.}
\institute{Vrije Universiteit Amsterdam\newline De Boelelaan 1105, 1081 HV Amsterdam, the Netherlands}
\maketitle
\def\thefootnote{*}\footnotetext{These authors contributed equally to this work.}\def\thefootnote{\arabic{footnote}}
\begin{abstract}
This study investigates the elicitation of empathy toward a third party through interaction with social agents. Participants engaged with either a physical robot or a voice-enabled chatbot, both driven by a large language model (LLM) programmed to exhibit either an empathetic tone or remain neutral. The interaction is focused on a fictional character, Katie Banks, who is in a challenging situation and in need of financial donations. The willingness to help Katie, measured by the number of hours participants were willing to volunteer, along with their perceptions of the agent, were assessed for 60 participants. Results indicate that neither robotic embodiment nor empathetic tone significantly influenced participants' willingness to volunteer. While the LLM effectively simulated human empathy, fostering genuine empathetic responses in participants proved challenging.

\end{abstract}
\section{Introduction}
\vspace{-0.5em}
Empathy plays a crucial role in human-human interaction (HHI) for building positive relationships and effectively communicating~\cite{fuller}. In human-computer interaction (HCI), mimicking human empathetic responses can enhance the quality of conversation and improve user satisfaction. Expressing empathy entails recognizing and validating the others' emotions, responding compassionately and supportively through active listening and understanding the user's affective state. The incorporation of empathetic behavior in virtual agents and social robots enhances their ability to interact successfully with humans. Empathetic agents are perceived as more relatable and trustworthy, leading to improved user satisfaction and engagement~\cite{bagheri2021}. Conversely, previous research has shown that empathy towards robots can be successfully elicited from humans~\cite{seo,frederiksen}. Moreover, Seo et al.~\cite{seo} pointed out that a physical robot may receive more empathy from humans compared to a non-physical one, highlighting the importance of embodiment.

Current-generation dialogue systems (DSs), on the other hand, exhibit distinctive features relevant to empathy, such as autonomy and usage of typical human language. This enriches interpersonal relationships and increases both credibility and engagement~\cite{chaves}. Unlike chatbots, physical robots provide cues like gestures and presence, which can impact user perceptions and emotional engagement. Comparing physical and virtual agents can reveal how embodiment influences empathy.

While most research has focused on empathy perceived from and directed toward robots, the use of social agents to elicit empathy from humans towards a third party remains largely unexplored. According to Park et all. and Bagheri et all.~\cite{park,bagheri2023} enhancing empathy towards a third party through social agent interaction could improve understanding and communication.

Therefore, the originality and novelty of our work lie in the fact that it centers on the incorporation of empathetic behavior in a social agent by employing an LLM to elicit empathy from a human towards a third person. The Katie Banks paradigm~\cite{coke}, which revolves around empathy toward a fictional character, is adopted to elicit empathy from humans. Thus, the main objective of the present work is to investigate in what manner robotic embodiment and empathetic tone (empathy vs. non-empathy) influence empathy elicited from humans towards a third party. By comparing a physical robot to a chatbot, we aim to discern the unique contributions of physical embodiment in eliciting empathetic responses. We hypothesize that a robotic embodiment is more successful in eliciting empathy from humans (H1) and  a robot that adopts an empathetic tone elicits more empathy from humans (H2).

\section{Related work}
\vspace{-0.75em}
Past work on human-robot interaction (HRI) and HCI has primarily focused on enhancing the quality of interactions by incorporating various human-like behaviors, including empathy. Niculescu et al.~\cite{niculescu} investigated the enhancement of HCI quality by manipulating voice pitch, expression of humor and display of empathy in a social robot. Compared to a non-empathetic robot, an empathetic one is perceived as more emotionally engaging. The findings suggest that an empathetic robot leads to improved interactions and user satisfaction in customer service scenarios, with users perceiving robot behavior to be appealing.

Another study by Bagheri et al.~\cite{bagheri2021} showcased a framework based on reinforcement learning for social robots that allows them to detect the user’s affective state and express empathy. The robot can provide participants with a sense of comfort and confidence, improving their enjoyment and well-being. Similarly, previous studies have demonstrated positive outcomes with the incorporation of empathy in HCI~\cite{leite,jesin,bickmore}. Therefore, a robot that expresses empathy is expected to evoke more empathy than a neutral robot from humans.

On the other hand, a negative user experience can be the result of empathy expression that does not match the user's emotional state~\cite{cramer,ochs}. To address this, Park and Wang~\cite{park} present a conceptual framework that considers elements for designing an empathetic robot. These elements include domain dependency, associated with the scenario in which a robot is employed and the extent to which empathy is expressed in a particular scenario, and nuanced modulation of empathy, which refers to giving different responses and varying strength of empathetic cues depending on the context. These considerations are applied by specifying that the agents' responses should vary in expressed empathy based on whether they are programmed to be empathetic or non-empathetic.

LLMs pave the way for dynamically generating empathetic responses due to their capability of instantly producing results from a prompt without extensive training and simplifying implementation through APIs for streamlined code integration. In addition, they require reduced computational resources and allow adaption to various language processing tasks through effective fine-tuning~\cite{kasneci}. Alessa and Al-Khalifa~\cite{alessa} showed that an LLM-based social chatbot is capable of having empathetic conversations. Another work by Loh and Raamkumar~\cite{loh} explored the production of empathetic responses by LLMs in the context of mental health care. Compared to conventional DSs, responses generated by LLMs were considered more empathetic. This implies that LLMs are a viable option for the implementation of empathetic social agents. The prompt from Loh and Raamkumar~\cite{loh}, which consisted of solely four sentences, is reused in our work.

Using a robotic embodiment, as opposed to a chatbot, enables the manipulation of the robot's voice. To compensate for this, the chatbot is equipped with a voice. Previous research findings suggest that physical presence and non-verbal cues significantly enhance user engagement and empathy~\cite{lee,wainer}. Moreover, an agent that adopts anthropomorphic characteristics yields more enjoyment and trust~\cite{klein}. Our work aims to build on these insights by integrating both LLM capabilities and robotic embodiment to maximize empathetic interactions. Unlike previous work, our research focuses on the combined effect of voice modulation and physical presence of social agents on elicited empathy rather than only textual or verbal empathy. We aim to provide a more comprehensive understanding of how different modalities contribute to elicited empathy, perceived empathy and user satisfaction. In the context of this study, elicited empathy refers to the empathy experienced by the user as a result of interacting with the social agent, whereas perceived empathy refers to the extent to which the recipient believes that the social agent is demonstrating empathy.

\section{Methodology}
\vspace{-0.5em}
The methodology involves a comprehensive approach, consisting of the interaction design, technical implementation and study design. The experiments involve a brief interaction between the participant and social agent, either Pepper or a chatbot. Both agents employ an LLM, prompted to hold a conversation regarding Katie Banks. The objective is to measure elicited empathy from the participants and their perception of the robotic agent they interacted with.

\subsection{Study design}
The primary objective of the current study is to investigate the elicited empathy towards a third party from participants who interact with a social agent. Specifically, the aim is to compare the levels of elicited empathy after interaction with either a physical robot or chatbot, both equipped with an LLM programmed to express empathy or to remain neutral in terms of empathy expression. The study adopts a between-subjects design and includes the following four experimental conditions:

\begin{itemize}\small
    \item Condition I: The physical robot displays empathetic expressions.
    \item Condition II: The physical robot refrains from empathy expressions.
    \item Condition III: The chatbot displays empathetic expressions.
    \item Condition IV: The chatbot refrains from empathy expressions.
\end{itemize}

\vspace{-1em}

\subsubsection{Participants}
60 participants (48 male, 11 female, 1 undisclosed) aged between 18 and 45 were recruited using convenience sampling. Each participant was randomly assigned with equal distribution across the experimental conditions, resulting in 15 participants per condition. The average age was 27.7, 27.5, 23.7 and 25.0 for the empathetic robot, non-empathetic robot, empathetic chatbot and non-empathetic chatbot, respectively. Prior to conducting the study, we ensured compliance with the ethical code of the Faculty of Science at Vrije Universiteit Amsterdam by performing an ethics self-check provided by the Research Ethics Review Committee Faculty of Science (BETHCIE).

\subsubsection{Procedure}
\begin{figure}
    \vspace{-1em}
    \centering
    \begin{minipage}{0.475\textwidth}
    \includegraphics[width=\textwidth]{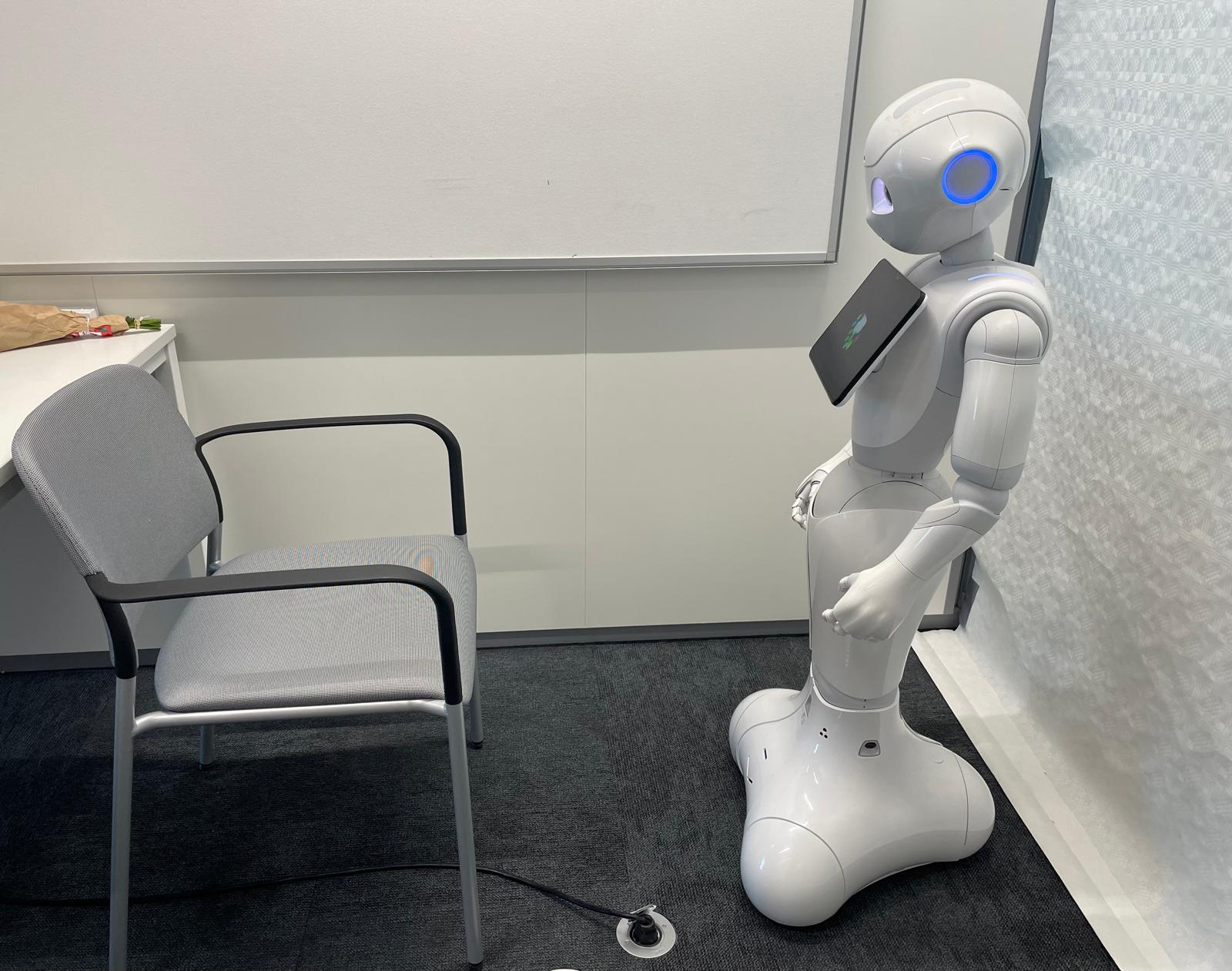}
    \vspace{-2em}
    \caption{Experimental setup with the robot, where the participant is seated in front of the robot during the interaction.}
    \label{fig:exp_setup}
  \end{minipage}
  \hfill
  \begin{minipage}{0.44\textwidth}
    \includegraphics[width=\textwidth]{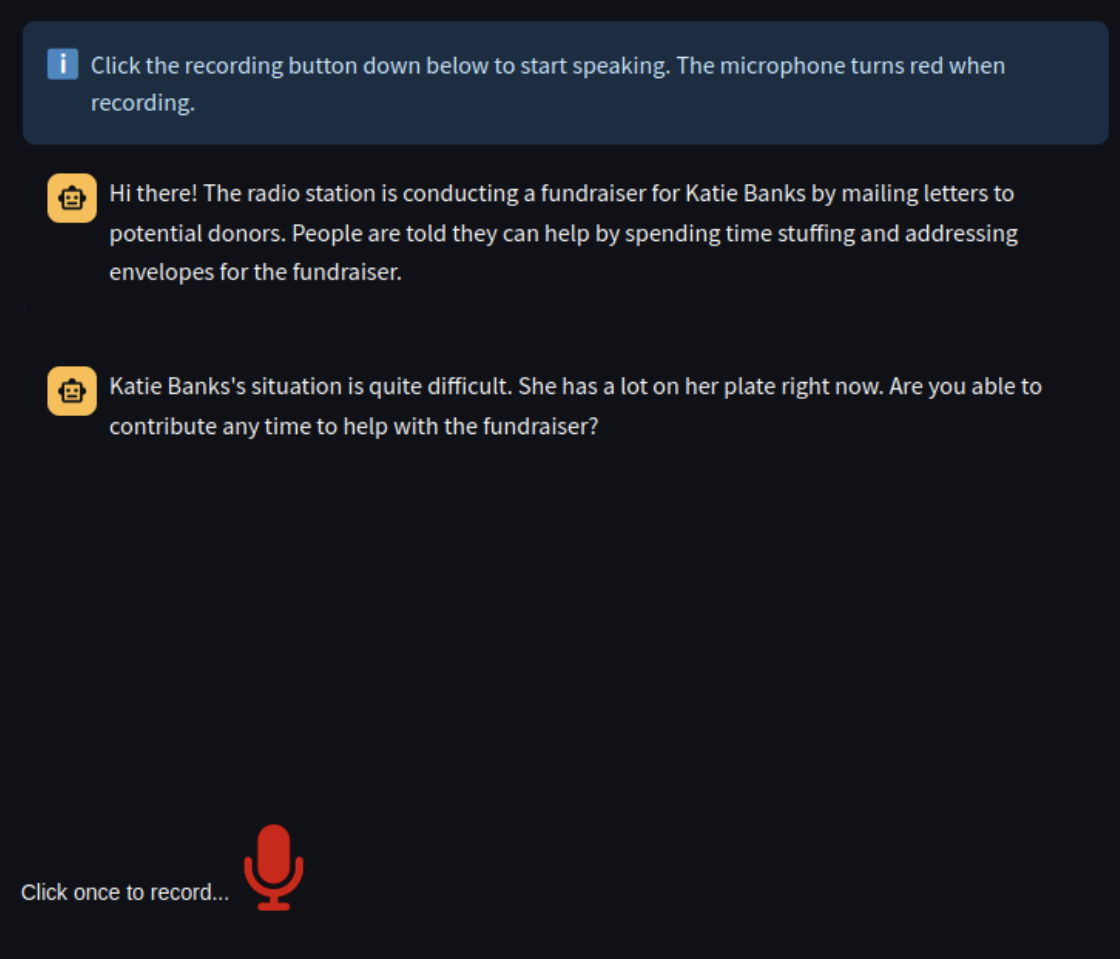}
    \vspace{-2em}
    \caption{User interface of the chatbot, showing the initial message to every participant and a response generated by the LLM.}
    \label{fig:chatbot_ui}
  \end{minipage}
\end{figure}

Participants engaged in a brief conversation with either the robot or chatbot. Refer to Fig.~\ref{fig:exp_setup} for the experimental setup with Pepper. See Fig.~\ref{fig:chatbot_ui} for the chatbot user interface. Prior to the interaction, the participants received instructions by voice and were allowed to request clarification. The participants first read an article about Katie Banks to familiarize themselves with her story. After reading the story, they engaged in a conversation with an agent, aiming to determine if they were willing to volunteer to help stuff envelopes for the fundraiser for Katie. Following the interaction, participants completed a questionnaire, which facilitated the measurement of participants' perceptions on the agents' empathy and their overall interaction experiences. The survey was composed of the State Empathy Scale~\cite{shen} and subscales from the Almere Model~\cite{heerink}. The former scale includes items that assess emotional and cognitive empathy, whereas the latter scale evaluates aspects related to the interaction quality. We used the subscales interaction, likability, perceived sociability, social presence, and trust from the Almere model~\cite{heerink}. All scales are Likert type and require participants to rate statements ranging from 1 (strongly disagree) to 5 (strongly agree). Participants also indicated the number of hours they are willing to donate to help Katie Banks, providing a quantitative measure of the influence of the agent's empathetic behavior. Next to the survey responses, user feedback and transcripts of the interactions were analyzed to investigate user perceptions. The provided answers were anonymized and cannot be linked to an identifiable entity.

\vspace{-2em}

\subsection{Interaction Design}
The Katie Banks paradigm, developed by Batson~\cite{coke}, is a well-established method used to elicit empathy through storytelling. Participants are first presented with a story about Katie Banks, a young woman facing significant personal challenges, including family issues and financial instability, in the form of a news article. After reading the article, participants hold a conversation with either Pepper or a chatbot, each programmed to display either empathetic or non-empathetic behavior. Following the interaction, participants complete a questionnaire to indicate their willingness to help Katie in terms of volunteering hours and to evaluate their perception regarding the interaction and empathy of the agent.

The interaction with the agent is initiated by the agent itself. The agent always starts the conversation by greeting the participant and informing them about a fundraiser for Katie Banks for which they can help stuff envelopes. A response related to Katie Banks is, then, provided by the agent. At the beginning of the interaction, the participant is informed that they can speak after hearing a beep from the robot. Once Pepper finishes speaking, a distinct beep signals the participant's turn to speak and voice recording begins. These auditory cues, designed with unique sounds, help establish clear communication signals and a structured turn-taking pattern during the interaction. Participants who interact with the chatbot instead sit in front of a laptop and press a virtual microphone button to speak. There is a silence detection mechanism in place for both Pepper and the chatbot that automatically stops recording after detecting four seconds of silence. It was found during our pilot study that this duration works effectively. Regarding ethical considerations, the recorded responses of the participants are not saved.

\subsection{Technical implementation}
\vspace{-0.5em}
For the interaction, Pepper as well a chatbot are used, both programmed to hold a conversation with the user about Katie Banks's story and their willingness to volunteer to help her. Participants interact with either an empathetic or a non-empathetic agent, which differ in empathy expression. Two different prompts were used for generating responses. One prompt instructed the LLM to generate responses with an empathetic tone, while the second prompt instructed the generation of responses with a neutral tone in terms of empathy. To this extent, the prompt from Loh and Raamkumaar~\cite{loh} was found to be effective for conversation. Their prompt was adopted and adjustments were made they were deemed necessary. The adjusted prompt includes the story of Katie Banks and information on a fundraiser that is being conducted to help Katie. In addition, the prompt specifies that the conversation should solely revolve around Katie Banks to avoid off-topic interactions, whereby the user is prompted to talk about Katie Banks's story as well as their willingness to volunteer to help Katie with a fundraiser by the agents. The non-empathetic prompt is similar, however only reuses the first sentence of the previously provided prompt. Instead of reusing the third sentence, instructions to stay neutral in terms of empathy are included.

\begin{figure}
  \centering
  \includegraphics[scale=0.4]{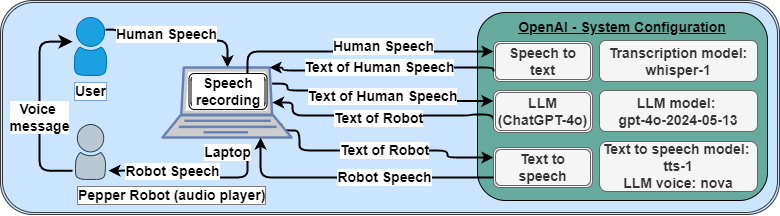}
  \vspace{-1em}
  \caption{System overview illustrating the interaction design.}
  \label{fig:system_overview}
\end{figure}

During the interaction, the user's vocal response is captured and converted to text, which is then processed by the OpenAI LLM API (GPT-4o) to generate a response. To ensure consistency, a laptop microphone is used for both the robot and chatbot. However, participants were unaware of its use during the robot experiments. Technical details for interactions with Pepper are shown in Fig.~\ref{fig:system_overview}. The chatbot has a similar design, using the laptop's audio system for voice output and the monitor for text display.

\vspace{-1em}

\vspace{-2em}

\section{Results and discussion}
\vspace{-0.5em}
Participants were asked to indicate the number of hours they would volunteer after interacting with the empathetic robot, non-empathetic robot, empathetic chatbot, and non-empathetic chatbot. The majority of participants (7–11) indicated a willingness to volunteer 2-4 hours across all agent types. Fewer participants (1–5) expressed interest in volunteering either 0 or 8-10 hours, with slight variations across agent types. Overall, the empathetic chatbot garnered the highest number of participants willing to volunteer 2-4 hours (11), while the non-empathetic robot had a relatively more even distribution across the different hour ranges.

\begin{figure}
    \centering
    \includegraphics[width=0.8\linewidth]{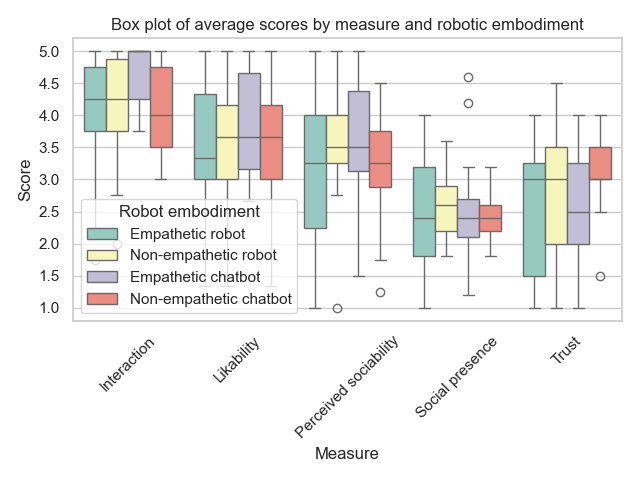}
    \vspace{-1em}
    \caption{Average scores for each selected subconstruct from the Almere model~\cite{heerink}, including interaction, likability, perceived sociability, social presence and trust, assigned by participants to agents from each experimental condition.}
    \label{fig:emp_percep_plot}
\end{figure}

The mean volunteering hours were 3.4 (SD = 2.97) for the empathetic robot, 4.0 (SD = 2.93) for the non-empathetic robot, 4.6 (SD = 2.03) for the empathetic chatbot, and 3.0 (SD = 3.00) for the non-empathetic chatbot. A one-way ANOVA revealed no statistically significant difference in volunteering hours across groups, (F(3, 26) = 0.341, p = .796). The average scores for the subjective measures collected for both type of agent are depicted in Fig.~\ref{fig:emp_percep_plot}.

\vspace{-1em}

\vspace{-1.5em}

One-way ANOVA tests revealed no statistically significant differences in interaction, likeability, perceived sociability, social presence, or trust between the different agent types (F(3, 26) = 1.59, p = .203; F(3, 26) = 0.47, p = .705; F(3, 26) = 0.52, p = .671; F(3, 26) = 0.16, p = .922; F(3, 26) = 1.16, p = .335). As a result, there is no evidence in favor of H1 and H2. Neither robotic embodiment nor empathetic tone appeared to be particularly effective in eliciting empathy from participants. This is in line with the feedback from the participants, where five out of 30 participants found the interaction with the robot pleasant and six out of 30 participants reported a similar experience with the chatbot.

Familiarity with the robot was recorded by asking participants whether they were familiar with Pepper. Considering solely those participants that interacted with Pepper, a t-test showed a significant association between familiarity and average interaction score. There was a significant difference in interaction scores of participants that were familiar (M = 4.66, SD = 1.083) and those not familiar with Pepper (M = 3.898, SD = 1.072); t(1) = 2.619, p = .017). This suggests that prior exposure to technology did influence the perceived quality of the interaction.

Concerning perceived empathy, mean average scores of 25.3, 25.7, 24.1 and 24.8 were obtained from participants assigned to the empathetic robot, non-empathetic robot, empathetic chatbot and non-empathetic chatbot, respectively. On average, the robot obtained higher scores regarding perceived empathy compared to the chatbot. Interestingly, both non-empathetic agents received a higher empathy perception score. Two participants perceived the agents as emotionless and ingenuine, whereas feedback from other participants indicated that empathetic agents were not perceived as such. Notwithstanding, one-way ANOVA test indicated no significant different between the average perceived empathy scores among the experimental conditions (F-value = .18 and p-value = .908).

The results align with Thellman et al. \cite{thellman2016physical}, showing no significant differences in how physical and virtual agents are perceived during interactions. Several factors may explain these null results. First, it is possible that the empathetic cues employed by the agents were too subtle or not sufficiently distinct to significantly affect the participants' responses. If the empathetic behaviors were not prominently displayed, participants may not have been influenced by them, leading to the observed lack of significant differences. Additionally, participants with prior experience or familiarity with the technology might have had preset expectations that overshadowed the specific behaviors exhibited during the interaction.

Since the chatbot allows pauses, incorporating similar pauses into the robot’s interactions could increase the quality of the interaction. Participants also indicated that quicker responses could enhance the interaction quality. In addition, repeated questions and conversations that cycled back to the same points caused annoyance among participants, suggesting that the conversation flow needs refinement.

Moreover, the LLM showed a tendency to hallucinate when asked unprompted questions, generating non-existent websites or locations. This presents a challenge for real-world applications where accuracy is essential, and a fallback approach for unanticipated questions could improve reliability. It is important to acknowledge that the sample size of 60 participants may have limited the study's statistical power, making it harder to detect subtle differences. A larger sample could improve sensitivity. 

\section{Conclusion}
\vspace{-0.5em}
This study investigated the elicitation of empathy through interactions with two types of agents: a physical robot and a chatbot, both employing an LLM that exhibited either empathetic or neutral behavior. Despite prior research suggesting that empathetic agents enhance user experience and satisfaction~\cite{bagheri2021,niculescu}, our findings did not reveal significant differences in the number of volunteering hours participants were willing to contribute, nor in their perceptions of empathy or interaction quality based on the type of agent.

These null results challenge earlier studies emphasizing the importance of physical embodiment and non-verbal cues in fostering empathy~\cite{lee,wainer,klein}. While such research highlights the potential of anthropomorphic robots and embodied agents to create more engaging and empathetic interactions, our findings suggest that empathy elicitation is a complex process that may not rely solely on the agent's form or its expressions of empathy.

Future research should explore alternative metrics for assessing empathy, such as monetary donations or acts of kindness, which may provide deeper insights. Additionally, the relatively small sample size of 60 participants limits the generalizability of our findings; larger samples would likely yield more robust results. Improving the interaction design of physical robots is also essential for maintaining consistency across experimental conditions.

Participants indicated that faster response times would enhance the human-likeness of interactions, suggesting that future systems should incorporate more dynamic response generation. Consistent with Loh and Raamkumar's findings~\cite{loh}, the LLM-generated responses were perceived as empathetic and engaging, reinforcing the potential for LLMs to contribute meaningfully to human-like interactions. Overall, this research underscores the importance of refining both the design and evaluation metrics of social agents to better understand their role in empathy elicitation and human-agent interaction.
\bibliographystyle{unsrt}
\bibliography{bibliography}
\end{document}